\documentstyle[preprint,aps,version2]{revtex}

\begin{document}

\textheight=185mm 
 
\thispagestyle{empty} 

\begin{title} 
Electroproduction of Transversely Polarized Vector Mesons\\ 
 Via A Quantum Mechanical Anomaly
\end{title}

\author{Pervez Hoodbhoy$^1$  and Wei Lu$^2$  }
\begin{instit}
$1$. Department of Physics,
Quaid-e-Azam University,
Islamabad 45320, Pakistan
\end{instit} 

\begin{instit} 
$2$. Department of Physics,
University of Maryland,
College Park, Maryland 20742, USA
\end{instit} 

\begin{abstract}
By explicit calculation we demonstrate that, in variance with the classical
prediction, the leading twist contribution to the exclusive
electroproduction of transversely polarized vector mesons from the nucleon
does not vanish beyond the leading order in perturbation theory. This
appears to be due to a quantum-mechanical anomaly, in the sense that a
classical symmetry of the field theory is broken by quantum corrections.
\end{abstract}

\centerline{ 
UMD PP\#99-087~~~~
DOE/ER/40762-175~~~~ 
February 1999}

\newpage
\setcounter{page}{1}
Recently, there has been considerable interest in the electroproduction of
vector mesons from a nucleon by a highly virtual longitudinally-polarized
photon \cite{ryskin,diehl}. Much of this interest follows the study of
off-forward parton distributions \cite{ji} in the nucleon, which have the
potential of providing insights into the nucleon's deep structure.
Among others, Collins, Frankfurt and Strikman \cite{collins}
showed that  the reaction amplitude
for the diffractive   meson electroproduction
can be factorized  into a form of convolutions
of the off-forward parton
distributions and  meson wave functions with hard
scattering coefficients.  Hence, it is desirable to 
study the feasibility to access the  off-forward parton
distributions in meson electroproduction processes.

Of particular interest is the electroproduction of transversely polarized
vector mesons because this may provide a handle to access two 
twist-2 chiral-odd off-forward parton distribution functions
$H_{T}(x,\xi )$ and $E_{T}(x,\xi )$. 
(We refer to Ref. \cite{hood} for a complete categorization
of the 6 quark and 6 gluon twist-two off-diagonal distribution functions 
 as well as  the definitions of $H_{T}(x, \xi)$
 and $E_{T}(x, \xi)$.) Experimentally,
chiral-odd distributions are notoriously difficult to measure because only
by matching with some other chiral-odd quantities can they make
non-vanishing contributions. As noted in Ref. \cite{collins}, the leading
twist wave function of transversely polarized vector mesons is chirally odd.
Thus there arises the possibility of accessing chiral-odd off-forward parton
distributions by studying the production of  transversely polarized vector
mesons.

Unfortunately, it turns out that the hard scattering coefficients associated
with matching chiral-odd off-forward parton distributions with chiral-odd
vector meson wave functions vanish \cite{man} at leading order in strong
coupling ($\alpha _{s}$). More interestingly, Diehl, Gousset and Pire 
\cite{diehl} presented a proof that these hard coefficients vanish to {\it all
orders} in perturbation theory. If this is true, it is both good and bad
news. The good news is that there will not be any leading twist chiral-odd
contaminations in the measurement of chiral-even distributions. The bad news
is that at leading twist it is impossible to access the chiral-odd parton
distributions by means of vector meson electroproduction.

In this Letter we demonstrate that quantum effects invalidate the proof of
vanishing hard coefficients for the transversely polarized vector meson 
electroproduction. By   explicit calculations at one loop level, 
we show that  those hard coefficients  do not generally vanish 
in perturbation theory.

To be specific, consider 
\[
\gamma ^{*}(q,e_{L})+N(P,S)\to V(K,e_{T})+N(P^{\prime },S^{\prime }), 
\]
where the first and second symbols in the parentheses stands for the
particle momentum and spin vector, respectively. As usual, we define the
average momentum and momentum difference for the initial- and
final-state nucleons: 
\begin{eqnarray}
&&\bar{P}=\frac{1}{2}(P+P^{\prime })\ , \\
&&\Delta =P^{\prime }-P\ .
\end{eqnarray}

It is most convenient to work in the frame in which $\bar{P}$ and $q$ are
collinear with each other and put them in the third direction. Since we are
going to deal with the light-cone dominated scattering processes, 
we  introduce two conjugate light-like vectors $p^{\mu }$ and $%
n^{\mu }$ in the third direction in the sense that $p^{2}=n^{2}=0$ and $%
p\cdot n=1$. Correspondingly, the relevant momenta can be parameterized as
follows: 
\begin{eqnarray}
&&q^{\mu }=-2\xi p^{\mu }+\nu n^{\mu }\ , \\
&&\bar{P}^{\mu }=p^{\mu }+\frac{\bar{M}^{2}}{2}n^{\mu }\ , \\
&&\Delta ^{\mu }=-2\xi (p^{\mu }-\frac{\bar{M}^{2}}{2}n^{\mu })+\Delta
_{\perp }^{\mu }\ ,
\end{eqnarray}
with $\nu =Q^{2}/(4\xi )$, $Q^{2}=-q^{2}$ and $\bar{M}^{2}=M^{2}-\Delta
^{2}/4$. In accord with this choice of coordinates, the longitudinal
polarization vector of the virtual photon reads 
\begin{equation}
e_{L}^{\mu }=\frac{1}{Q}\left( 2\xi p^{\mu }+\nu n^{\nu }\right) \ .
\end{equation}
At lowest twist we can safely  approximate   the particle momenta as follows:
\begin{eqnarray}
&&P^{\mu }=(1+\xi )p^{\mu }+\cdots \ , \\
&&P^{\prime \mu }=(1-\xi )p^{\mu }+\cdots \ , \\
&&K^{\mu }=\nu n^{\mu }+\cdots \ .
\end{eqnarray}

The basic idea of factorization for vector meson electroproduction is
illustrated in Fig. 1. According to the factorization theorem, the dominant
mechanism is a single quark scattering process. The reaction amplitude is
approximated as a product of three components: the hard partonic scattering,
the non-perturbative matrix associated with the nucleon, and the matrix
associated with the vector meson production. The active quark has to come
back into the nucleon blob after experiencing the hard scattering. On the
nucleon side, the initial and final quarks can be thought of carrying
momenta $(x+\xi )p$ and $(x-\xi )p$, respectively.

By decomposing the nucleon matrix one has, 
\begin{eqnarray}
&&\int \frac{d\lambda }{2\pi }e^{i\lambda x}\langle P^{\prime }S^{\prime }|
\bar{\psi}_{\alpha ,i}(-\frac{1}{2}\lambda n)\psi _{\beta ,j}(\frac{1}{2}
\lambda n)|PS\rangle  \nonumber \\
&&~~~~~~~~=\frac{\delta _{ij}\sigma _{\alpha \beta }^{\rho \lambda }}{4N_{c}}
\left[ H_{T}(x,\xi )\bar{U}(P^{\prime }S^{\prime })\sigma _{\rho \lambda
}U(PS)+E_{T}(x,\xi )\bar{U}(P^{\prime }S^{\prime })\frac{\gamma _{[\rho
}i\Delta _{\lambda ]}}{M}U(PS)\right] +\cdots \ .
\end{eqnarray}
To save space,  we displayed here only the twist-2 chiral-odd
off-forward parton distributions that we will focus on. 
In the above, $\alpha $ and $\beta $ are 
the quark  spinor indices, $i$ and $j$ the color
indices, $N_{c}=3$ is the number of quark colors. By $[\rho \lambda ]$ we
mean antisymmetrization of the two indices. The ellipses represent all other
distribution functions irrelevant for the forthcoming discussion. For
simplicity, it has been assumed that there is only one flavor of quark and
correspondingly  the flavor index is suppressed. 
Also suppressed is the gauge link operator in the definition of 
the matrix elements.

On the side of the vector meson production, the momenta that the quark and
antiquark carry can be approximately parameterized as $z\nu n$ and $(1-z)\nu
n$. Similarly, one can write down the following decomposition for the
non-perturbative matrix associated with the vector meson production, 
\begin{equation}
\int {\frac{d\lambda }{2\pi }}e^{-i\lambda z}\langle 0|\bar{\psi}_{\beta
,i}(-\frac{1}{2}\lambda \bar{n})\psi _{\alpha ,j}(\frac{1}{2}\lambda \bar{n}
)|K,e_{T}\rangle =\frac{\delta _{ij}\sigma _{\beta \alpha }
^{\rho \lambda }}
{2\sqrt{N_{c}}}F_{T}(z)K_{\rho }e_{T\lambda }^{*}+...\ ,
\end{equation}
where $F_{T}(z)$ is a twist-two chiral-odd vector meson wave function.

The possibility to access chiral-odd quantities arises
as  $H_{T}(x,\xi )$ and $E_{T}(x,\xi )$  are matched 
with $F_{T}(z)$ in the closed active quark loop. 
Without loss of generality, we may put the amplitude 
into the following form: 
\begin{eqnarray}
{\cal A} &=&\left( \frac{e}{Q}\right) \int dxdz\frac{F_{T}(z)}{(x-\xi
+i\epsilon )(1-z)}\sum_{{\rm diagrams}}C_{i}f_{i}(x,\xi ,z)  \nonumber \\
&&\times \left[ H_{T}(x,\xi )\bar{U}(P^{\prime }S^{\prime })\rlap/n\rlap/%
e_{T}U(PS)+E_{T}(x,\xi )\frac{e_{T}\cdot \Delta }{M}\bar{U}(P^{\prime
}S^{\prime })\rlap/nU(PS)\right] \ ,
\end{eqnarray}
where $C_{i}$ is the color factor and $f_{i}(x,\xi ,z)$  the corresponding
kinematical factor. Our task is to calculate $\sum C_{i}f_{i}(x,\xi ,z)$ to
next-to-leading order in the strong coupling constant.

At the tree level, there are two Feynman diagrams for the hard partonic
scattering, as shown in Fig. 2. This corresponds to the fact that either
before or after it is struck by the virtual photon, the active quark must
undergo a hard scattering to adjust its momentum so as to  form the
final-state vector meson. (Remember that in our chosen frame, both initial-
and final-state nucleons move in the third plus direction, while the
vector meson goes in the opposite direction.) By working in the Feynman
gauge, one can most easily understand why the hard coefficients $f_{i}(x,\xi
,z)$ vanish at tree level. Actually, it is a direct
consequence of the following {\it 4-dimensional} identity: 
\begin{equation}
\gamma ^{\mu }\sigma ^{\rho \lambda }\gamma _{\mu }=0.  \label{sigma}
\end{equation}
Here the $\sigma ^{\rho \lambda }$-matrix comes either from the density
matrix associated with the off-forward quark helicity-flip distribution in
the nucleon (see Fig. 2a) or from that associated with the chiral-odd
light-cone wave function of the vector meson (See Fig. 2b), while the two 
$\gamma $-matrices sandwiching $\sigma ^{\rho \lambda }$ correspond to the
hard gluon scattering exchange. If one sticks to the 4-dimensions, it can be
generally shown that the hard coefficients vanish to {\it all orders} in
perturbation theory. We must emphasize, however, 
 that Eq. (\ref{sigma}) holds
only in the 4-dimensions. If one goes beyond the leading order, loops will
necessarily occur and lead to divergences that must be regulated. It is most
advantageous to use dimensional regularization for both ultraviolet and
infrared divergences. If one works in $(4-2\varepsilon )$ dimensions, 
$\gamma ^{\mu }\sigma ^{\rho \lambda }\gamma _{\mu }$  
will be of ${\cal O}(\varepsilon )$ 
and the Dirac trace for the quark loop will no longer be
zero. Hence, by canceling the poles from the loop integrations against the 
$\varepsilon $'s arising from the Dirac traces, 
some non-vanishing terms may survive. 
As will become clear, both ultraviolet and infrared divergences can
make contributions to  the non-vanishing coefficients at one-loop level.

For convenience, our calculations are done in the Feynman gauge. We found
out that it is preferable to group the diagrams by their color structure. We
work with renormalized perturbation theory, so all the self-energy and
vertex corrections are understood to be accompanied by the corresponding
ultraviolet counter-terms.

The diagrams shown in Fig. 3 are characteristic of the three-gluon vertex
and possess a common color factor of $C_{{\rm fig.3}}=(N_{c}^{2}-1)/(4\sqrt{%
N_{c}})$. The ultraviolet divergences in those vertex corrections are
understood to have been canceled by their counter-terms, so only the infrared
divergences make contributions. For individual diagrams, there are 
\begin{eqnarray}
&&f_{{\rm 3a}}=-4\alpha _{s}^{2}\left[ 1+\frac{1}{2z}\log (1-z)+\frac{\xi }{%
\xi +x}\log \frac{\xi -x}{2\xi }\right] \ , \\
&&f_{{\rm 3b}}=\alpha _{s}^{2}\left[ -2-\frac{2}{z}\log (1-z)\right] \ , \\
&&f_{{\rm 3c}}=\alpha _{s}^{2}\left[ -2-\frac{4\xi }{\xi +x}\log \frac{\xi -x%
}{2\xi }\right] \ , \\
&&f_{{\rm 3d}}=-4\alpha _{s}^{2}\ , \\
&&f_{{\rm 3e}}=-4\alpha _{s}^{2}\ .
\end{eqnarray}
Summing over all the five diagrams in Fig. 3, one has 
\begin{equation}
\sum\limits_{{\rm fig.3}}C_{i}f_{i}(\xi ,x,z)=-\frac{N_{c}^{2}-1}{\sqrt{N_{c}%
}}\alpha _{s}^{2}\left[ 4+\frac{1}{z}\log (1-z)+\frac{2\xi }{\xi +x}\log 
\frac{\xi -x}{2\xi }\right] \ .  \label{1}
\end{equation}

Fig. 4 contains a group of diagrams that have a common color factor but do
not contribute to the hard coefficients. The first three drop out simply
because their Dirac traces vanish even in the $(4-2 \varepsilon)$ space. 
The last two do not contribute because their vertex corrections contain no
infrared divergences, while the ultraviolet divergences are canceled by the
counter-terms.

Shown in Fig. 5 are another group of diagrams that have the same color
factor $C_{{\rm fig.~5}}=-(N_{c}^{2}-1)/(4N_{c}^{2}\sqrt{N_{c}})$. Some
diagrams in this group require lengthy calculation. After considerable
algebra, we obtain the contributions for  individual diagrams as follows: 
\begin{eqnarray}
&&f_{{\rm 5a}}=\alpha _{s}^{2}\left\{ -\frac{2}{\varepsilon _{I}}-4+2\ln
\left[ \frac{(1-z)(\xi -x)}{2\xi }\frac{Q^{2}e^{\gamma }}{4\pi \mu ^{2}}%
\right] \right\} \ , \\
&&f_{{\rm 5b}}=\alpha _{s}^{2}\left\{ -\frac{2}{\varepsilon _{I}}-4+2\ln
\left[ \frac{(1-z)(\xi -x)}{2\xi }\frac{Q^{2}e^{\gamma }}{4\pi \mu ^{2}}%
\right] \right\} \ , \\
&&f_{{\rm 5c}}=\alpha _{s}^{2}\left\{ -2-2\frac{1-z}{z}\ln (1-z)\right\} \ ,
\\
&&f_{{\rm 5d}}=\alpha _{s}^{2}\left\{ -2-2\frac{\xi -x}{\xi +x}\ln \frac{\xi
-x}{2\xi }\right\} \ , \\
&&f_{{\rm 5e}}=\alpha _{s}^{2}\left\{ -\frac{2}{\varepsilon _{I}}-\frac{2}{z}%
\log (1-z)-2\log \left[ \frac{z^{2}(\xi -x)}{2\xi }\frac{Q^{2}e^{\gamma }}{%
4\pi \mu ^{2}}\right] \right\} \ , \\
&&f_{{\rm 5f}}=\alpha _{s}^{2}\left\{ -\frac{2}{\varepsilon _{I}}-\frac{4\xi 
}{\xi +x}\log \frac{\xi -x}{2\xi }-2\log \left[ \frac{(1-z)(\xi -x)^{2}}{%
(2\xi )^{2}}\frac{Q^{2}e^{\gamma }}{4\pi \mu ^{2}}\right] \right\} \ , \\
&&f_{{\rm 5g}}=\alpha _{s}^{2}\left\{ +\frac{2}{\varepsilon _{I}}-2+2\frac{%
1-z}{z}\log (1-z)+2\log \left[ \frac{z(\xi +x)}{2\xi }\frac{Q^{2}e^{\gamma }%
}{4\pi \mu ^{2}}\right] \right\} \ , \\
&&f_{{\rm 5h}}=\alpha _{s}^{2}\left\{ +\frac{2}{\varepsilon _{I}}-2+2\frac{%
\xi -x}{x+\xi }\log \frac{\xi -x}{2\xi }+2\log \left[ \frac{z(\xi +x)}{2\xi }%
\frac{Q^{2}e^{\gamma }}{4\pi \mu ^{2}}\right] \right\} \ , \\
&&f_{{\rm 5i}}=\alpha _{s}^{2}\left\{ -\frac{4}{\varepsilon _{I}}-2\frac{1-z%
}{z}\log (1-z)-2\frac{\xi -x}{x+\xi }\log \frac{\xi -x}{2\xi }
 -4\log \left[ \frac{(1-z)(\xi -x)}{2\xi }\frac{%
Q^{2}e^{\gamma }}{4\pi \mu ^{2}}\right] \right\} \ ,
\end{eqnarray}
where  $1/\varepsilon_I$ is the infrared pole, 
$\mu^2$ the scale parameter in the dimensional renormalization, 
and $\gamma$ the Euler constant. Summing over all the diagrams 
in Fig. 5, we have, 
\begin{equation}
\sum\limits_{{\rm fig.5}}C_{i}~f_{i}(\xi ,x,z)=\frac{N_{c}^{2}-1}{N_{c}^{2}%
\sqrt{N_{c}}}\alpha _{s}^{2}\left[ 4+\frac{1}{z}\log (1-z)+\frac{2\xi }{\xi
+x}\log \frac{\xi -x}{2\xi }\right] \ .  \label{2}
\end{equation}

At this stage, we comment on the one-loop self-energy corrections for the
hard scattering partonic processes. Since we work with the renormalized
perturbation theory, we need not consider the self-energy insertions either
on the incoming or outgoing quark lines. Instead, we need to recalculate the
tree diagrams, shown in Fig. 2, in $(4-2\varepsilon )$-dimensions and
include a factor of $\sqrt{Z_{F}}$ for each external quark line of the hard
scattering part. The ultraviolet pole in $Z_{F}$ can be compensated by the $%
\varepsilon $ factor from the tree-level trace in $(4-2\varepsilon )$
dimensions. This is exactly where the ultraviolet divergences make their
contribution. As a result, we have 
\begin{equation}
\sum_{{\rm tree}}C_{i}~f_{i}(\xi ,x,z)=\frac{(N_{c}^{2}-1)^{2}}{4N_{c}^{2}%
\sqrt{N_{c}}}\times 4\alpha _{s}^{2}.  \label{3}
\end{equation}
On the other hand, diagrams with a self-energy 
insertion onto   an internal line do not 
contribute because they have no infrared divergences.

At this stage, we have exhausted all the one-loop diagrams for the hard
scattering process. Combining Eqs. (\ref{1}), (\ref{2}) and (\ref{3}), we
finally reach the following compact expression for the one-loop
coefficients: 
\begin{equation}
\sum\limits_{{\rm NTL}}C_{i}f_{i}(x,\xi ,z)=-4\alpha _{s}^{2}\frac{C_{F}^{2}%
}{\sqrt{N_{c}}}\left[ 4+\frac{1}{z}\log (1-z)+\frac{2\xi }{\xi +x}\log \frac{%
\xi -x}{2\xi }\right] \ 
\end{equation}
with $C_{F}=(N_{c}^{2}-1)/(2N_{c})$, which is the central result of this 
Letter.

In summary, our explicit one-loop calculations demonstrate that, within the
context of the factorization theorem, the leading twist part of the
amplitude for the transversely polarized  
vector meson electroproduction does not vanish, except at
the tree level. The fact that the hard scattering coefficients are finite is
directly a consequence of the need to regularize and renormalize
perturbation theory. In turn, this is a consequence of the infinite number
of degrees of freedom in a field theory. Therefore,  it is a kind of 
quantum-mechanical anomaly in the sense that the  classical symmetry of
field theory is broken by quantum corrections.

\medskip

{\bf Acknowledgments }

We would like to thank Xiangdong Ji for valuable suggestions and criticisms.
PH thanks the Pakistan Science Foundation for support. This work is
supported in part by funds provided by the U.S. Department of Energy
(D.O.E.) under cooperative agreement DOE-FG02-93ER-40762.

\newpage 

\centerline{\large Caption} 

Fig. 1.  Illustration of factorization of amplitude for
vector meson electroproduction.

Fig. 2.  Tree-level hard partonic scattering diagrams. 

Fig. 3. One-loop corrections to the hard partonic scattering with 
a three-gluon vertex. 

Fig. 4.  A group of one-loop diagrams that  share a color factor but do 
not make non-vanishing contributions to the hard scattering coefficients. 

Fig. 5.  A group of one-loop diagrams that share a color factor and make 
non-vanishing contributions to the hard scattering coefficients. 

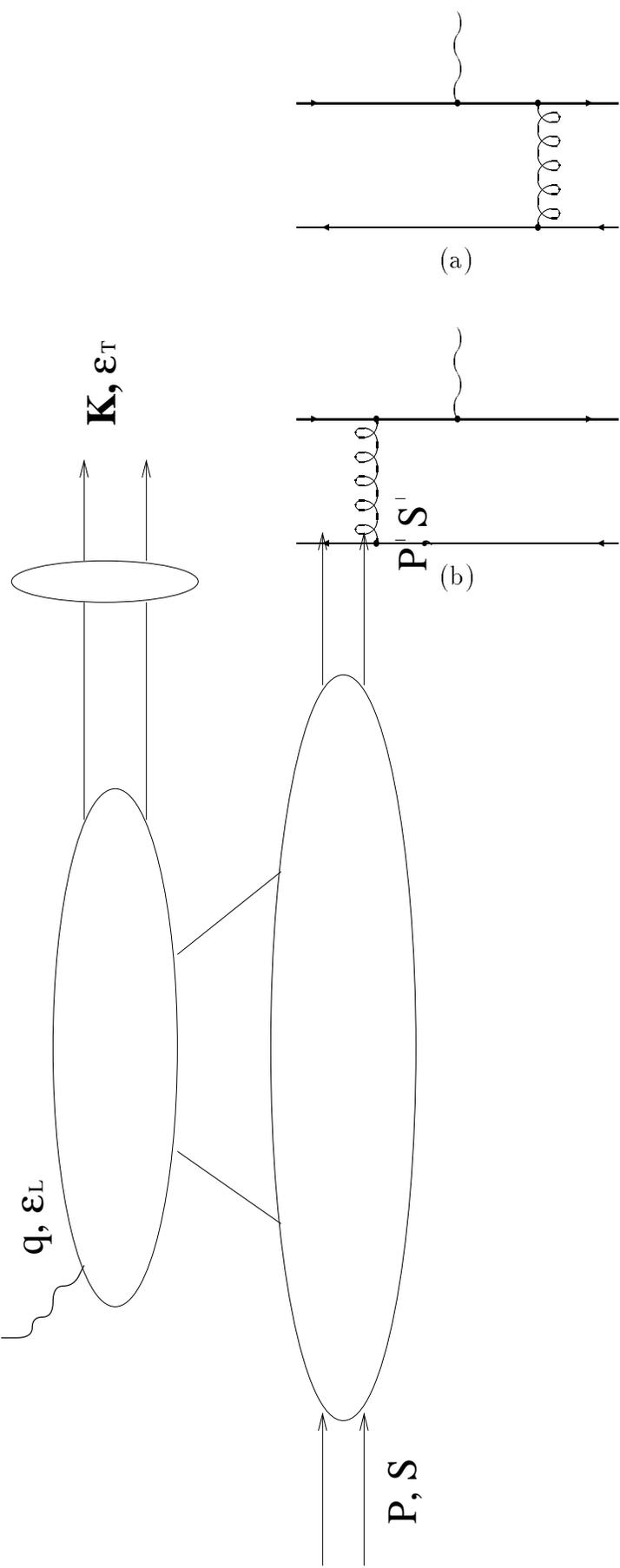
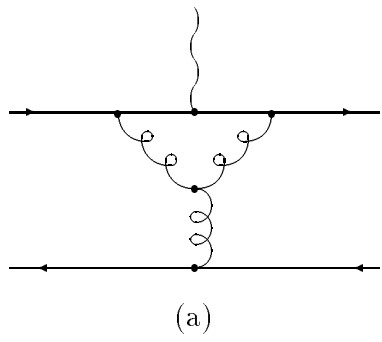
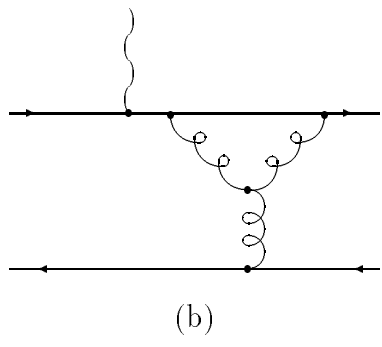
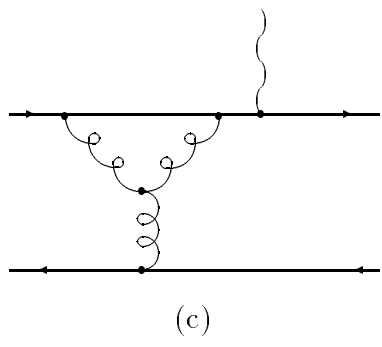
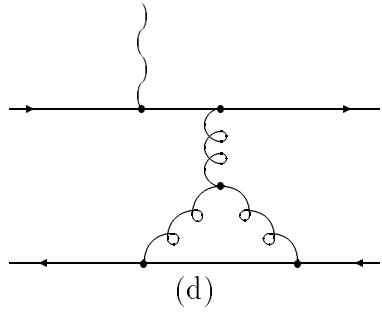
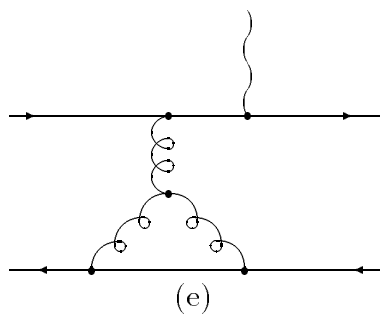
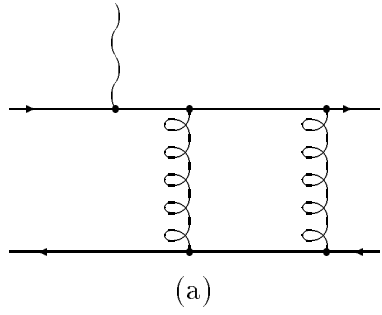
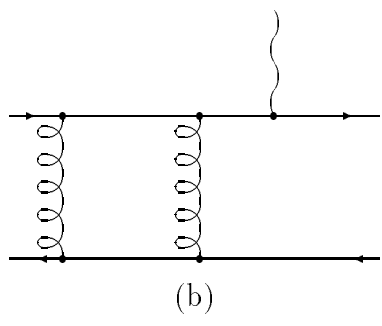
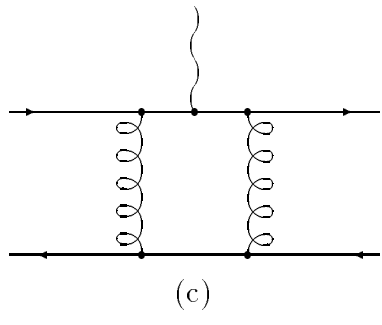
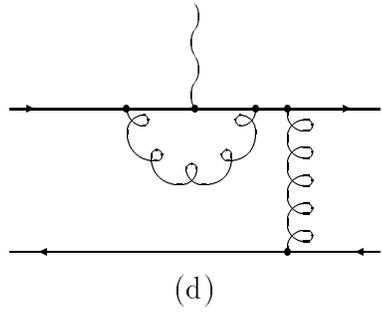
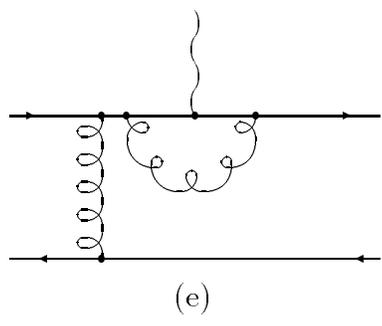
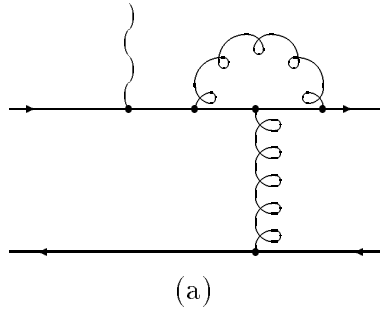
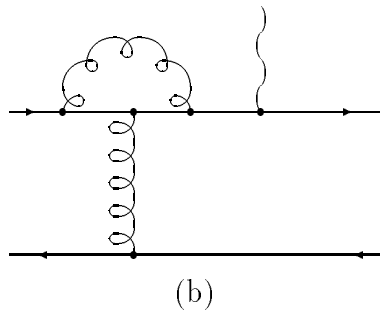
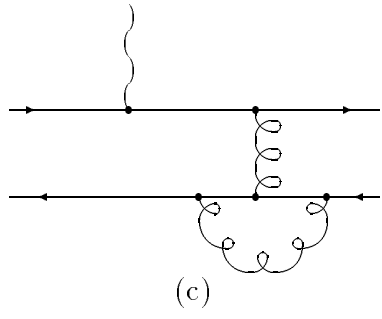
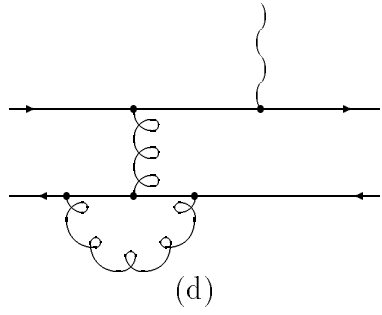
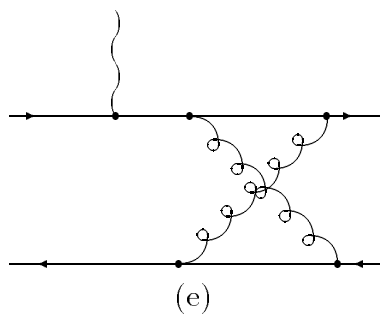
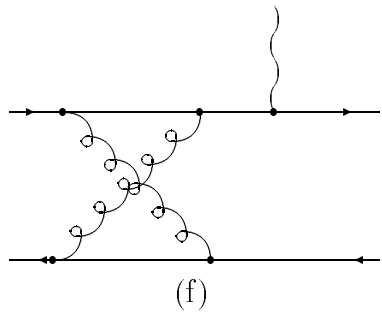
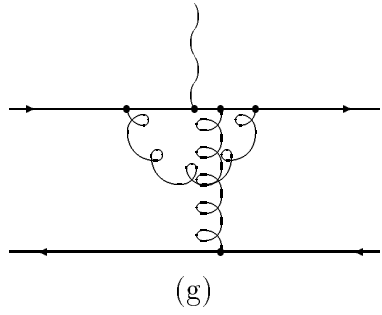
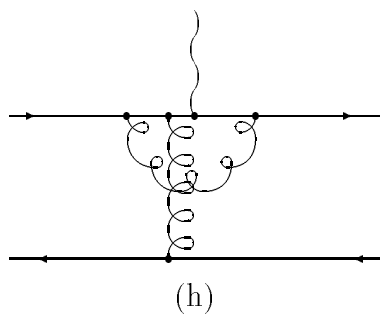
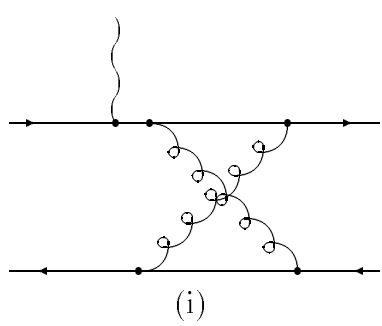
\end{document}